\documentclass[12pt, a4paper]{article}
\usepackage{amsmath,amssymb,amsfonts}
\usepackage{graphicx}
\usepackage[colorlinks]{hyperref}
\usepackage{makecell}
\usepackage{cite}
\setcellgapes{3pt}

\begin{document}

\title{Impact of space-time curvature coupling on the vacuum energy induced by a magnetic topological defect in flat space-time of arbitrary dimension}%

\author{V.M. Gorkavenko$^{1,2}$\thanks{Corresponding author. \textit{Email address:} \textbf{gorkavol@knu.ua} (Volodymyr Gorkavenko)}, O.V. Barabash$^{1}$, I.V. Ivanchenko$^{1}$, P.O.~Nakaznyi$^{3}$,\\M.S. Tsarenkova$^{1}$, N.S. Yakovenko$^{1}$, A.O. Zaporozhchenko$^{1}$ \\
\phantom{jvjjv}\\
\it \small ${}^1$ Faculty of Physics, Taras Shevchenko National University of Kyiv,\\
\it \small 64, Volodymyrs'ka str., Kyiv 01601, Ukraine\\
\it \small ${}^2$ Bogolyubov Institute for Theoretical Physics, National Academy of Sciences of Ukraine,\\
\it \small 14-b, Metrolohichna str., Kyiv 03143, Ukraine\\
\it \small ${}^3$ Institute of Physics and Technology, Igor Sikorsky Kyiv Polytechnic Institute\\ 
\it \small 37, prospect Beresteiskyi, Kyiv 03056, Ukraine}

\date{}

\maketitle

\begin{abstract}
We have investigated vacuum polarization of a quantized charged massive scalar field in the presence of a magnetic topological defect, modeled as an impenetrable tube of finite thickness carrying magnetic flux. At the tube's surface, we imposed a general Robin boundary condition. Our analysis demonstrates that, in flat space-time, the total induced vacuum energy is independent of the coupling 
$\xi$ of the scalar field's interaction with the space-time curvature only in the special cases of Dirichlet and Neumann boundary conditions. For general Robin boundary conditions, however, the total induced vacuum energy depends on the coupling $\xi$ in a flat space-time and exhibits a nontrivial dependence on the parameter of the Robin boundary condition. We investigated the dependence of this effect not only on Robin's boundary condition parameter, but also on the tube thickness and the space-time dimensionality. We conclude that careful measurements of vacuum polarization effects in flat space-time may, in principle, provide an independent way to probe the $\xi$ coupling.

Keywords: vacuum polarization; topological defect; Aharonov-Bohm effect;  Casimir effect.
\end{abstract}

\maketitle

\section{Introduction}

After the Casimir paper \cite{Casimir:1948dh}, it became clear that the presence of external boundaries leads to changes in the vacuum energy density. The observed effect of changes in vacuum energy density is the Casimir force. 
This problem was considered for different boundaries' shapes and materials, see, e.g., \cite{Elizalde:1995hck,Mostbib,Bordag:2001qi,Klimchitskaya:2023niz}.

In this paper, we study vacuum polarization by a magnetic topological defect, modeled as an infinitely long cylindrical tube that is impenetrable to a matter field and contains a magnetic flux inside. In such a configuration, the Aharonov-Bohm effect \cite{Aharonov:1959fk} plays a crucial role, as the enclosed magnetic field can influence the quantum matter field located outside the tube.  {By vacuum polarization we mean the modification of the quantum vacuum caused by an external field or boundary, which alters the spectrum of virtual particles and leads to changes in observable quantities.} Within the framework of second-quantized field theory, such a topological defect gives rise to an induced vacuum energy, vacuum current and the corresponding magnetic flux in the exterior region (see, e.g., the pioneering studies carried out in the singular magnetic vortex approximation \cite{Serebryanyi:1985blr,Gornicki:1990kq,Flekkoy:1990pn,Parwani:1991bc}). The imposed boundary condition at the tube’s surface has a significant impact on the behavior of the matter field outside. Since the observed quantum effects originate from both the boundary condition and the internal magnetic flux, this phenomenon is commonly referred to as the Casimir–Bohm–Aharonov effect \cite{Sitenko:1997zf}.

The phenomenon of vacuum polarization near the linear magnetic topological defects has a wide range of physical applications.
These topological defects can be derived from phase transitions with spontaneous gauge symmetry breaking in the early Universe or condensed matter physics. The impenetrable magnetic tube model can be applied to describe magnetic cosmic strings, the formation of which was
possible in the early Universe \cite{Kibble:1980mv,Vilenkin:1981kz,Vilenkin:2000jqa,Hindmarsh:1994re}. In condensed matter physics, the impenetrable magnetic tube can be considered as a model of Abrikosov-Nielsen-Olesen vortex in type-II superconductors \cite{Abrikosov:1956sx,Nielsen:1973cs} or as disclinations in nanoconical structures of two-dimensional materials like graphene, see, e.g., \cite{Krishnan:1997cs,heibergandersen2006carbon,Sitenko:2007ewb,naess2009carbon,Sitenko:2018sag,Sitenko:2019oul,Barkeshli:2025cjs}.

In the general case, one can also consider the interaction of the scalar matter field ($\psi$) with the space-time curvature \cite{penrose1964relativity,chernikov1968quantum,Callan:1970ze,Birrell:1982ix}
\begin{equation}\label{0}
\mathcal{L}=({\mbox{$\nabla$}}_\mu\psi)^*({\mbox{$\nabla$}}^\mu\psi)-(m^2+\xi R)\psi^*\psi,
\end{equation}
where ${\mbox{ $\nabla$}}_\mu$ is the covariant derivative involving both affine and
bundle connections and $m$
is the mass of the scalar field, $R^{\mu\nu}$ is the Ricci tensor, $R=g_{\mu\nu}R^{\mu\nu}$ is the
scalar curvature of space-time and $\xi$ is the coupling  of the
scalar field to the scalar curvature of space-time.
The energy-momentum tensor (EMT) in this case can be written as 
\begin{equation}\label{a1}
   T^{\mu\nu}=T^{\mu\nu}_{can}+2\xi\left(g^{\mu\nu}\Box-\nabla^\mu\nabla^\nu-R^{\mu\nu}\right)
   \psi^*\psi\,,
\end{equation}
where
\begin{equation}\label{a2}
   T^{\mu\nu}_{can}=\nabla^\mu\psi^*\nabla^\nu\psi+
 \nabla^\nu\psi^* \nabla^\mu\psi-g^{\mu\nu}\mathcal{L}.
\end{equation}
Here, $T^{\mu\nu}$ is obtained by varying the action with respect to the metric, $T^{\mu\nu}_{can}$ corresponds to the canonical EMT from Noether's theorem.  As one can see, the dependence of the EMT of the scalar field (and in particular the energy density $T^{00}$) on the coupling $\xi$ of the scalar field's interaction with the space-time curvature remains even in the case of flat space-time.  However, an interesting question is whether the total energy of the system can depend on the coupling $\xi$?

The coupling $\xi$ plays a key role in a wide range of physical models, including early Universe inflation \cite{Faraoni:2000wk,Kaiser:2015usz}, dark matter and large-scale structure formation \cite{Sankharva:2021spi}. It also affects cosmic microwave background anisotropies, baryon acoustic oscillations, the matter power spectrum \cite{Geng:2015nnb}, and even the stability of our Universe \cite{Branchina:2019tyy}.  Among all possible values for the parameter $\xi$, there are several special values. The case of $\xi=0$ is called the case of minimal coupling.
In the case of $(d+1)$-dimensional space-time, the conformal invariance of the theory is achieved at 
\begin{equation}
\xi=\xi_c=\frac{d-1}{4d}.
\end{equation}
In the case of $(3+1)$-dimensional space-time $\xi_c=1/6$. It was shown in \cite{Sitenko:2002zp} that the value $\xi = 1/4$ also plays an important  {technical} role, which we will discuss further.

In this paper, we will investigate the vacuum polarization of a charged massive scalar matter field $\psi$ in the flat space-time of arbitrary dimension in the background of a magnetic tube of finite transverse size 
$r_0$, which is impenetrable to the matter field. We will be interested in the dependence of the induced vacuum energy on the coupling to the space-time curvature. On the surface of the tube, we consider  {general} Robin-type boundary conditions
\begin{equation}\label{Robin}
    (\cos \theta\, \psi + \sin \theta\, r \partial_r \psi)|_{r_0} =0,
\end{equation}
where  {$r_0$ is the radius of the tube} and $\theta$ is a parameter of the boundary condition, $-\pi/2 \leq \theta< \pi/2$.
The special cases of 
$\theta=0$ and $\theta=-\pi/2$ correspond to Dirichlet and Neumann boundary conditions, respectively. For these special cases, the induced vacuum energy has been studied in \cite{Gorkavenko:2009qn,Gorkavenko:2011qg,Gorkavenko:2013rsa} (with the Dirichlet boundary condition) and in \cite{Gorkavenko:2022xtv} (with the Neumann boundary condition).

In \cite{Gorkavenko:2024vuy} it was shown that, in $(2+1)$-dimensional flat space-time with a fixed tube radius $mr_0 = 1/100$, the total vacuum energy of a scalar field induced in the background of an impenetrable tube with Robin-type boundary conditions and magnetic field inside depends on the coupling $\xi$ between the scalar field and space-time curvature.
 
There is something magical about the fact that the coupling $\xi$ of the scalar field with the curvature of space-time can, in principle, manifest itself in observable global quantities (induced energy) by carefully measuring the effects of vacuum polarization in the background of linear topological defects in flat space-time. So, in this paper, we will continue the research of \cite{Gorkavenko:2024vuy}. We consider in detail the dependence of the total induced vacuum energy of the scalar field induced by the impenetrable tube with a magnetic field inside with the Robin-type boundary condition on its surface on the transverse size of the tube $mr_0$ and the coupling $\xi$ in flat space-time of a physically interesting case of dimension $(3+1)$ and arbitrary dimension as well.

\section{Induced vacuum energy density. General relations}

We will use the Lagrangian \eqref{0} for a complex scalar massive field $\psi$  in
$(d+1)$-dimensional space-time. In the case of a
static background \cite{Birrell:1982ix,Grib:1980aih,Parker:2009uva}, the operator of the quantized charged scalar field,  {whose classical version appears in  Lagrangian \eqref{0}}, has the form
\begin{equation}\label{a11}
\Psi(x^0,\textbf{x})=\sum\hspace{-1.4em}\int\limits_{\lambda}\frac1{\sqrt{2E_{\lambda}}}
\left[e^{-iE_{\lambda}x^0}\psi_{\lambda}(\textbf{x})\,a_{\lambda}+
  e^{iE_{\lambda}x^0} \psi_\lambda^\ast(\textbf{x})\,b^\dag_{\lambda}\right],
\end{equation}
where $a^\dag_\lambda$ and $a_\lambda$ ($b^\dag_\lambda$ and
$b_\lambda$) are the scalar particle (antiparticle) creation and
annihilation operators; $\lambda$ is
the set of parameters (quantum numbers) specifying the state;
  $E_\lambda=E_{-\lambda}>0$ is the energy of the state; symbol
  $\sum\hspace{-1em}\int\limits_\lambda$ denotes summation over discrete and
  integration (with a certain measure) over continuous values of
  $\lambda$; wave functions $\psi_\lambda(\textbf{x})$ are the
  solutions to the stationary equation of motion  {that follow from Lagrangian \eqref{0}},
\begin{equation}\label{a12}
 \left\{-{\mbox{\boldmath $\nabla$}}^2  + (m^2+\xi R)\right\}  \psi_\lambda(\textbf{x})=E^2_\lambda\psi(\textbf{x}),
\end{equation}
where $\mbox{\boldmath $\nabla$ }$  is the covariant differential operator
in an external (background) field.

The vacuum energy density is determined as the vacuum
expectation value of the $T^{00}$ component of the EMT, which is given as \cite{Sitenko:2002zp,Gorkavenko:2024vuy}
\begin{equation}\label{a14}
\varepsilon=\langle \rm
vac|\left[\partial_0\Psi^\dag\partial_0\Psi+\partial_0\Psi\partial_0\Psi^\dag+(1/4-\xi)\mbox{\boldmath
$\nabla$}^2(\Psi^\dag\Psi+\Psi\Psi^\dag)\right]- \xi R^{00}(\Psi^\dag\Psi+\Psi\Psi^\dag)|\rm vac\rangle,
\end{equation}
 {where the vacuum state $|\mathrm{vac}\rangle$ is defined in the usual way by the conditions $a_\lambda|\rm vac\rangle=b_\lambda |\rm vac\rangle=0$.}
We now focus on the case of flat space-time, where $R^{\mu\nu}=0$, $R=0$. In this case, the vacuum energy density still depends on the coupling $\xi$ and can be written as
\begin{equation}\label{a14a}
\varepsilon=\varepsilon_{can}+(1/4-\xi)\varepsilon_\xi=\sum\hspace{-1.4em}\int\limits_{\lambda}E_\lambda\psi^*_\lambda(\textbf{x})\,\psi_\lambda(\textbf{x})+(1/4-\xi)\mbox{\boldmath
 $\nabla$}^2
   \sum\hspace{-1.4em}\int\limits_{\lambda}E^{-1}_\lambda\psi^*_\lambda(\textbf{x})\,\psi_\lambda(\textbf{x}),
\end{equation}
 where $\varepsilon_{can}$ denotes the canonical energy density $T^{00}_{can}$, which is independent of $\xi$. As one can see, the canonical relation for the energy density is restored at $\xi=1/4$ \cite{Sitenko:2002zp}.

   {Although linear topological defects can produce, in vacuum, a conical spacetime with a deficit angle, in this paper we focus on the simplified case without an angle deficit, and treat the region outside the magnetic tube as Minkowski space. This approximation is well justified because,
 even for cosmic strings, the deficit angle is expected to be small, $\Delta \phi=8 \pi \c{G} \mu< 10^{-6}$ \cite{Charnock:2016nzm},
where $\c{G}$ is Newton’s gravitational constant and $\mu$
 is the linear mass density of the string core.  
 Such a small deficit angle has a negligible influence on vacuum polarization effects near the tube even in higher-dimensional spaces, in contrast to the substantial angle deficits $\Delta \phi\sim 1$ that play an essential role in describing quantum phenomena associated with disclinations in nanoconical structures of two-dimensional materials; see, e.g., \cite{Sitenko:2007ewb,Sitenko:2019oul,Sitenko:2022gha}.}
 
We consider a static background described by a cylindrically symmetric tube of finite transverse size carrying a magnetic flux. The coordinate system is chosen such that the tube is oriented along the $z$-axis. In $(d+1)$-dimensional space-time, this configuration can be naturally generalized to 
 the  {$(d-1)$-tube} by adding an additional $d-3$  longitudinal directions.  {Thus, in $2+1$ space-time, we have a ring lying in the plane perpendicular to the 
$z$-axis; in $3+1$  space-time, we obtain a tube extending along the $z$-axis;
and in $4+1$ space-time, this generalises to a hypersurface consisting of all points that satisfy $x^2+y^2=r_0^2$ at arbitrary values of $z$ and additional $x^1$ coordinates.}
 The covariant derivative is defined as $\nabla_0=\partial_0$, $\mbox{\boldmath
$\nabla$}=\mbox{\boldmath $\partial$}-{\rm i} \tilde e\, {\bf V}$
where $\tilde e$ is the coupling of dimension
$m^{(3-d)/2}$, and the vector potential has only one nonvanishing component, 
\begin{equation}\label{4}
V_\varphi=\Phi/2\pi,
\end{equation}
outside the tube. Here, $\Phi$ denotes the gauge flux confined within the tube, and $\varphi$ is the angular coordinate in the polar system $(r, \varphi)$ on the plane transverse to the tube.
  
In flat space-time, the solution to Eq.\eqref{a12} that satisfies the Robin boundary conditions \eqref{Robin} outside the impenetrable tube of radius $r_0$ has the form
\begin{equation}\label{6}
\psi_{kn{\bf p}}({\bf x})=(2\pi)^{(1-d)/2}e^{{\rm i}\bf{p
x}_{d-2}}e^{{\rm i}n\varphi}\Omega_{|n- {\tilde e}
\Phi/2\pi|}(\theta,kr,kr_0),
\end{equation}
where $\theta$ is the parameter characterizing the boundary condition,
\begin{equation}\label{7}
\Omega_\rho(\theta,u,v)=\sin\mu_\rho(\theta,v) J_{\rho}(u)-\cos\mu_\rho(\theta,v) Y_{\rho}(u),
\end{equation}
\begin{align}
    & \sin\mu_\rho(\theta,v)=\frac{\cos\theta\, Y_{\rho}(v)+\sin\theta\, v Y'_{\rho}(v)}{\sqrt{\left[\cos\theta J_{\rho}(v)+\sin\theta\, v J'_{\rho}(v)\right]^2+\left[\cos\theta\, Y_{\rho}(v)+\sin\theta\, v Y'_{\rho}(v)\right]^2}},\\
  &  \cos\mu_\rho(\theta,v)=\frac{\cos\theta\, J_{\rho}(v)+\sin\theta\, v J'_{\rho}(v)}{\sqrt{\left[\cos\theta J_{\rho}(v)+\sin\theta\, v J'_{\rho}(v)\right]^2+\left[\cos\theta\, Y_{\rho}(v)+\sin\theta\, v Y'_{\rho}(v)\right]^2}},
\end{align}
$ 0 < k < \infty$,  {$\bf{p}=(p^z,p^1,...,p^{d-3})$, $\bf{x}_{d-2}=(z,x^1,...,x^{d-3})$, $ -\infty < p^z < \infty $,} $ -\infty < p^j < \infty $ for $ j = 1, \dots, d{-}3 $; and $ n \in \mathbb{Z} $, where $ \mathbb{Z}$ denotes the set of integer numbers. The functions $ J_\rho(u) $ and $ Y_\rho(u) $ are the Bessel functions of the first and second kinds of order $ \rho $, respectively.  A prime on a function denotes differentiation with respect to its argument.
The solutions given by Eq.~\eqref{6} satisfy the orthonormalization condition
\begin{equation}\label{eq8}
\int\limits_{r>r_0} d^{\,d}{\bf x}\, \psi_{kn{\bf p}}^*({\bf
x})\psi_{k'n'{\bf p}'}({\bf
x})= \frac{\delta(k-k')}{k}\,\delta_{n,n'}\,\delta^{d-2}(\bf{p}-\bf{p}').
\end{equation}

Unfortunately, the Exp.\eqref{a14a} exhibits ultraviolet divergences and requires renormalization, see, e.g., \cite{Mostbib}.  {In our setting, where the magnetic field is excluded from the region accessible to the matter field, the renormalization procedure reduces to a single subtraction: removing the contribution corresponding to the absence of magnetic flux, see \cite{Babansky:1999re}. Consequently, the induced vacuum energy is defined as the change in vacuum energy between an impenetrable tube carrying a magnetic flux and the same impenetrable tube without flux. Of course, both quantities also contain a surface divergence arising from the presence of the boundary (the tube surface) - the Casimir surface energy. This contribution is determined solely by the geometry/topology of the boundary and by the boundary conditions imposed on the fields  \cite{Mostbib,Bordag:2001qi,Bordag:2001ta,Vassilevich:2003xt}; it is independent of the presence of the magnetic flux 
inside the tube. Consequently, this divergence cancels upon subtraction and does not require the introduction of additional counterterms.}

Using \eqref{a14a} and \eqref{6}, along with the renormalization procedure described above, the renormalized vacuum energy density in $(d+1)$-dimensional space-time can be written as \cite{Gorkavenko:2013rsa}
\begin{equation}\label{c2qq}
\varepsilon_{ren}=(2\pi)^{1-d} \int\limits_{-\infty}^\infty d^{d-2} {\bf
p}\int\limits_0^\infty
  dk\,k\left(\sqrt{ {\bf p}^2+k^2+m^2}+\frac{1/4-\xi}{\sqrt{{\bf p}^2+k^2+m^2}}\triangle_r\right)G(\theta,kr,kr_0,\Phi),
\end{equation}
where  
$\triangle_r = \partial_r^2 + r^{-1} \partial_r$  
is the radial part of the Laplacian operator on the plane transverse to the $(d-2)$-dimensional tube.
 Here, the $G$ function is defined as
\begin{equation}\label{Gfunction}
    G(\theta,kr,kr_0,\Phi)=S(\theta,kr,kr_0,\Phi)-S(\theta,kr,kr_0,0),
\end{equation}
\begin{equation}\label{a29a}
S(\theta,kr,kr_0,\Phi)=\sum_{n\in\mathbb
 Z}\Omega^2_{|n- {\tilde e}
\Phi/2\pi|}(\theta,kr,kr_0).
\end{equation}

As can be seen from \eqref{c2qq}, the value 
$\xi=1/4$ is technically distinguished: it reduces the energy density to its canonical expression. The factor $(1/4-\xi)$ separates the terms in the renormalized energy according to their convergence behavior at large integration momenta. Furthermore, note that the induced energy density in $(d+1)$-dimensional space-time does not depend on additional spatial coordinates.

As a result of the infinite range of summation, the function $S$ and consequently the function 
$G$, as well as the induced quantities, depend only on the fractional magnetic flux  inside the tube
\begin{equation}\label{a29a1}
   F=\frac{\tilde e\Phi}{2\pi}-\left[\!\left[\frac{\tilde e\Phi}{2\pi}\right]\!\right],\quad(0\leq F < 1),
\end{equation}
where $[[u]]$ is the integer part of quantity $u$ (less than or equal to $u$). Moreover, there is a symmetry under the substitution $F \rightarrow 1-F$.
 The properties and various representations of the $S$ function are discussed in detail in \cite{Gorkavenko:2024vuy}. We will not dwell on them in the present paper.

\section{Induced vacuum energy in $2+1$ space-time}

In this section, we focus on $(2+1)$-dimensional space-time. In this case, the renormalized vacuum energy density takes the form
\begin{equation}\label{c2}
\varepsilon_{ren}=\frac1{2\pi}\int\limits_0^\infty
  dk\,k\left(  \sqrt{k^2+m^2}+\frac{1/4-\xi}{\sqrt{k^2+m^2}} \triangle_r\right)G(\theta,kr,kr_0, {F}),
\end{equation}
where $\triangle_r=\partial^2_r+r^{-1}\partial_r$ is the transverse radial part of the Laplacian.  {We have also redefined the function $G$, and in what follows, we will consider it to be expressed in terms of $F$ as specified in \eqref{a29a1}.}

For the subsequent numerical computations, it is convenient to rewrite \eqref{c2} in a dimensionless form \cite{Gorkavenko:2011qg}
\begin{equation}\label{c3}
r^3\varepsilon_{ren}=\alpha_+(\theta,\lambda,x_0,F)+(1/4-\xi)r^3\triangle_r\left(\frac{\alpha_-(\theta,\lambda,x_0,F)}{r}\right),
\end{equation}
where
\begin{equation}\label{c3ab}
\alpha_\mp(\theta,\lambda,x_0,F)=\frac{1}{2\pi}\int\limits_0^\infty
dz\,z\left[z^2+\left(\frac{x_0}\lambda\right)^2\right]^{\mp1/2}
G(\theta,z,\lambda z,F),
\end{equation}
and $x_0=mr_0$, $\lambda=r_0/r$, $\lambda\in[0,1]$.
 It is also useful to consider the dimensionless function of parameters $x=mr$ and $x_0=mr_0$ 
\begin{multline}\label{m2}
\tilde
\alpha_-(\theta,x,x_0,F)=r^3\triangle_r\left(\frac{\alpha_-(\theta,x,x_0,F)}{r}\right)=\\ \alpha_-(\theta,x,x_0,F)-x\frac{\partial
\alpha_-(\theta,x,x_0,F)}{\partial x}+x^2\frac{\partial^2
\alpha_-(\theta,x,x_0,F)}{\partial x^2},
\end{multline}
and then write the dimensionless induced vacuum energy density as
\begin{equation}\label{m3}
r^3\varepsilon_{ren}=\alpha_+(\theta,x,x_0,F)+(1/4-\xi)\tilde\alpha_-(\theta,x,x_0,F).
\end{equation}

Unfortunately, due to the complicated form of the one-particle solutions in the case of a finite-thickness magnetic tube \eqref{6}, the vacuum energy density cannot be computed analytically and must be evaluated numerically.  
In what follows, we restrict our consideration to the case of a half-integer magnetic flux, $F = 1/2$, for which the vacuum polarization effect is known to be maximal in the presence of a singular magnetic vortex~\cite{Sitenko:1997zf,Sitenko:2002zp}.
 The procedure and technical details of the numerical computations are presented in \cite{Gorkavenko:2009qn,Gorkavenko:2011qg,Gorkavenko:2024vuy}. 
The explicit form (plots) of the computed functions $\alpha_\mp$, $\tilde
\alpha_-$ and dimensionless induced vacuum energy density $r^3\varepsilon_{ren}$ and different values of coupling $\xi$ are presented in \cite{Gorkavenko:2024vuy}.

\begin{figure}[t!]
    \centering
    \includegraphics[width=0.8\textwidth]{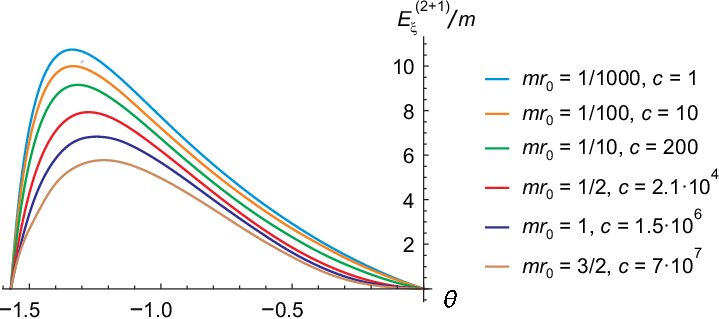}
    \caption{The total induced dimensionless vacuum energy $E_\xi^{2+1}/m$ \eqref{m5} in $(2+1)$-dimensional space-time as the function of parameter $\theta$ of Robin boundary conditions on the edge of the magnetic impenetrable tube is presented for different thicknesses of the tube. 
    For convenience, the above functions are multiplied by the coefficient $c$.  {For example, the value of the function presented in the figure for $mr_0=1/10$ is multiplied by a factor of $c=200$.}}
    \label{Fig1}
\end{figure}

The total vacuum energy induced by the impenetrable finite thickness magnetic tube is given by \vspace{-0.5em}
\begin{multline}\label{m4}
E^{(2+1)}=\int\limits_{0}^{2\pi}d\varphi\int\limits_{r_0}^\infty
\varepsilon_{ren} \,r dr=E^{(2+1)}_{can}+(1/4-\xi)E^{(2+1)}_\xi=\\=2\pi m \int\limits_{x_0}^\infty \frac{dx}{x^2}\left[ 
\alpha_+(\theta,x,x_0,F) +(1/4-\xi)\tilde\alpha_-(\theta,x,x_0,F)\right],\vspace{-1em}
\end{multline}
where $E^{(2+1)}_{can}$ corresponds to the canonical definition of $T^{00}$ component of the EMT \eqref{a2}.

Compared to the $\alpha_-$ and $\tilde\alpha_-$ functions, the computation of the $\alpha_+$ function requires considerably more computational effort, owing to the different degree of the factor $(z^2 + (mr)^2)$ in \eqref{c3ab}.  This means that it is quite difficult to compute the total vacuum energy induced by a magnetic tube with Roben's boundary condition on its surface. However, it is relatively easy to analyze the component $E_\xi$, which determines the dependence of the total induced vacuum energy on the coupling $\xi$. 
To do it, we can use integration by parts
\begin{equation}\label{m5}
E^{(2+1)}_\xi=2\pi m\int\limits_{x_0}^\infty \frac{\tilde\alpha_-(\theta,x,x_0,F)}{x^2}\,
dx=-2 \pi m x\left.\frac{\partial}{\partial x}\frac{\alpha_-(\theta,x,x_0,F)}{x}\right|_{x=x_0}.
\end{equation}

It was shown in previous papers that $E_\xi$ is zero 
in the case of the Dirichlet \cite{Gorkavenko:2011qg} and the Neumann boundary condition \cite{Gorkavenko:2024vuy} on the tube surface, but the induced vacuum energy $E_\xi$ is non-zero in the general case of the Robin boundary conditions, at least for negative values of the boundary condition
parameter $-\pi/2<\theta<0$.

The result of the numerical computation of $E_\xi^{(2+1)}$ is presented in Fig.\ref{Fig1} for different values of the tube thickness. As one can see, induced vacuum energy $E_\xi$ is zero at the boundaries of the interval $-\pi/2<\theta<0$  and is a positive function within this interval. As one can see, the vacuum effects increase strongly as the tube radius decreases. 
It is also worth noting that as the thickness of the tube decreases, the value of the boundary condition parameter at which the function reaches its maximum shifts toward the Neumann boundary condition, $\theta = -\pi/2$.

The case of positive values of the boundary condition parameter $\theta$ requires special attention. In this case, bound-state solutions of the Fock–Klein–Gordon equation contribute to vacuum polarization, leading to nontrivial vacuum effects, similar to those observed for the induced magnetic flux \cite{Sitenko:2022gha}.

\section{Induced vacuum energy in higher-dimensional spaces}\label{Sec4}

As shown in the previous sections, the induced vacuum energy density depends solely on the distance from the string within the plane transverse to the tube. Accordingly, in higher-dimensional space-times, the total induced energy becomes divergent due to integration  {of the induced energy density \eqref{c2qq}} over  {coordinates} perpendicular to this transverse plane. Therefore, it is meaningful to consider the induced energy only within the plane transverse to the tube.

As it was shown in \cite{Gorkavenko:2013rsa}, the case of the vacuum polarization in the presence of an impenetrable magnetic tube for the space-time of arbitrary dimension can be generalized from the case of $(2+1)$-dimensional space-time for the case of the Dirichlet boundary condition on the edge of the magnetic impenetrable tube, when $E^{(2+1)}_\xi(\theta=0)=0$.  In fact, due to changing the order of integration over $r$ and $\bf p$
an expression was obtained for induced vacuum energy $E^{(d+1)}_{can}$ within the plane transverse to the tube in $(d+1)$-dimensional space-time  via $E^{(2+1)}_{can}$, namely \vspace{-0.5em}
\begin{multline}\label{4s4}
E^{(d+1)}_{can}=(2\pi)^{2-d}\int\limits_{r_0}^\infty dr \,r \int\limits_{-\infty}^\infty d^{d-2}{\bf
p}\int\limits_0^\infty
  dk\,k\sqrt{{\bf p}^2+k^2+m^2}\, G(\theta,kr,kr_0,F)=\\
  =\frac2{r_0^{d-1}}\frac{(4\pi)^{1-d/2}}{\Gamma\left(\frac{d-2}2\right)}\int\limits_{x_0}^\infty
dv\, v^2 \left(v^2-x_0^2\right)^{\frac{d-4}2}\,
\mathcal{D}_{can}\left(F,\theta,v\right),
\end{multline}
where $\mathcal{D}_{can}$ defines induced energy in $(2+1)$-dimensional space-time
\begin{equation}\label{dop1}
E^{(2+1)}_{can}= m \mathcal{D}_{can}(F,\theta,mr_0),
\end{equation}
\begin{equation}\label{dddop2}
\mathcal{D}_{can}(F,\theta,y)=\int\limits_{y}^\infty\frac{dx}{x^2}\int\limits_0^\infty
dz\, z
\sqrt{z^2+x^2}\,G(\theta,z,z y/x,F).
\end{equation}

 \begin{figure}[t!]
    \centering
    \includegraphics{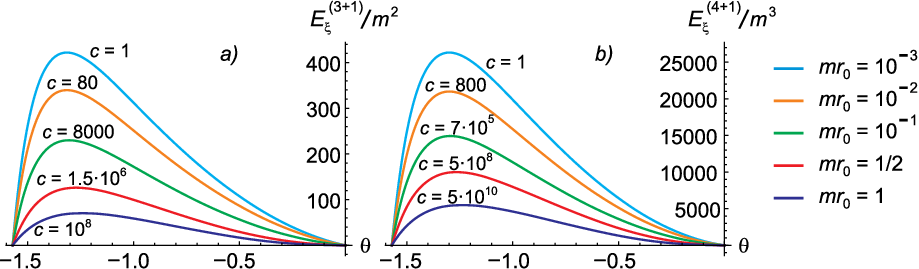}
    \caption{The induced dimensionless vacuum energy $E_\xi^{d+1}/m^{d-1}$ \eqref{4s5} within the plane transverse to the tube in $(d+1)$-dimensional space-time as the function of parameter $\theta$ of Robin boundary conditions on the edge of the magnetic impenetrable tube is presented for different thicknesses of the tube: \textit{a}) $d=3$, \textit{b}) $d=4$. For convenience, the above functions are multiplied by the coefficient $c$.  {For example, the value of the function presented in the figure \textit{a}) for $mr_0=10^{-1}$ is multiplied by a factor of $c=8000$.}}
    \label{Fig3}
\end{figure}

In this section, we will again be interested in only a part of the induced vacuum energy (within the plane transverse to the tube) proportional to $(1/4-\xi)$.  It can be shown, see Appendix A, that the corresponding energy can be written as  \vspace{-0.5em}
\begin{multline}\label{4s5}
E^{(d+1)}_{\xi}=(2\pi)^{2-d}\int\limits_{r_0}^\infty dr \,r \int\limits_{-\infty}^\infty d^{d-2}{\bf
p}\int\limits_0^\infty
  dk\,k\,\frac{\Delta_r G(\theta,kr,kr_0,F)}{\sqrt{{\bf p}^2+k^2+m^2}}=\\
  =\frac2{r_0^{d-1}}\frac{(4\pi)^{1-d/2}}{\Gamma\left(\frac{d-2}2\right)}\int\limits_{x_0}^\infty
dv\, v^2 \left(v^2-x_0^2\right)^{\frac{d-4}2}\,\mathcal{D}_\xi(F,\theta,v),
\end{multline}
where $\mathcal{D}_\xi$ defines induced energy in $(2+1)$-dimensional space-time
\begin{equation}\label{dop3}
E^{(2+1)}_{\xi}= m \mathcal{D}_{\xi}(F,\theta,mr_0),
\end{equation}
\begin{equation}\label{dddop4}
\mathcal{D}_{\xi}(F,\theta,y)=\int\limits_{y}^\infty dx\,x \Delta_x \int\limits_0^\infty
dz\, z\frac{G(\theta,z,z y/x,F)}{x\sqrt{z^2+x^2}},
\end{equation}
where $\triangle_r=\partial^2_x+x^{-1}\partial_x$.

 Comparing relations for $E^{(d+1)}_{can}$ \eqref{4s4} and $E^{(d+1)}_\xi$ \eqref{4s5}, one can see that they have a similar structure. It means that we can write general useful relations for the induced vacuum energy within the plane transverse to the tube in the space-time of dimension $d+1$ in the form
\begin{multline}\label{dddop5}
    E^{(d+1)}=E^{(d+1)}_{can}+(1/4-\xi)E^{(d+1)}_{\xi}
    =\\ =\frac2{r_0^{d-1}}\frac{(4\pi)^{1-d/2}}{\Gamma\left(\frac{d-2}2\right)}\int\limits_{x_0}^\infty
dv\, v^2 \left(v^2-x_0^2\right)^{\frac{d-4}2} \left[
\mathcal{D}_{can}\left(F,\theta,v\right)+(1/4-\xi)\mathcal{D}_{\xi}\left(F,\theta,v\right)\right]
\end{multline}
where $\mathcal{D}_{can}$ and $\mathcal{D}_{\xi}$ are given by \eqref{dddop2} and \eqref{dddop4}.

The results of our computations for the function $E^{(d+1)}_\xi$ \eqref{4s5} are shown in Fig.\ref{Fig3} for the representative cases $d=3$ and $d=4$  {, and half-integer magnetic flux $F = 1/2$. 
One can see that the total energy \eqref{dddop5}, in addition to the canonical contribution, contains a term $(1/4-\xi)\,E^{(d+1)}_{\xi}$, and the quantity $E^{(d+1)}_{\xi}$ does not vanish in spaces of arbitrary dimension. The value of $E^{(d+1)}_{\xi}$ is determined by the parameter $\theta$ of the general Robin boundary condition and exhibits a similar dependence on this parameter across different space–time dimensions. Specifically, the induced energy $E^{(d+1)}_{\xi}$ vanishes at the endpoints of the interval $-\pi/2 \leq \theta \leq 0$, while inside this interval it is a smooth function with a single maximum. }
One can also see that the induced vacuum dimensionless energy in $d+1$ space-time $E^{(d+1)}_\xi/m^{d-1}$ is rapidly increasing with increasing space dimension $d$ for sufficiently small transverse sizes of the topological defect $mr_0$  {and rapidly increasing as $mr_0$ decreases}.
 The main question is: how large are the values of the $E^{(d+1)}_\xi/m^{d-1}$ functions? To answer this, let's compare them with the values of the $E^{(d+1)}_{can}/m^{d-1}$ functions.

As already noted, computing the function $E^{(d+1)}_{can}$ for different values of the boundary condition parameter $\theta$ is a highly cumbersome task. In fact, this was done only for two cases, namely the Dirichlet \cite{Gorkavenko:2009qn,Gorkavenko:2011qg,Gorkavenko:2013rsa} and Neumann boundary \cite{Gorkavenko:2022xtv} conditions. Nevertheless, qualitative conclusions and estimates can still be made. Based on the results of \cite{Sitenko:2022gha} (see Figs.~3 and 4 therein), we infer that for Robin boundary conditions with $-\pi/2 < \theta < 0$, the induced vacuum energy $E^{(d+1)}_{can}$ lies between the minimal value for the Dirichlet boundary condition $(\theta = 0)$ and the maximal value for the Neumann boundary condition $(\theta = -\pi/2)$.  

\begin{figure}[t!]
    \centering
    \includegraphics[width=\textwidth]{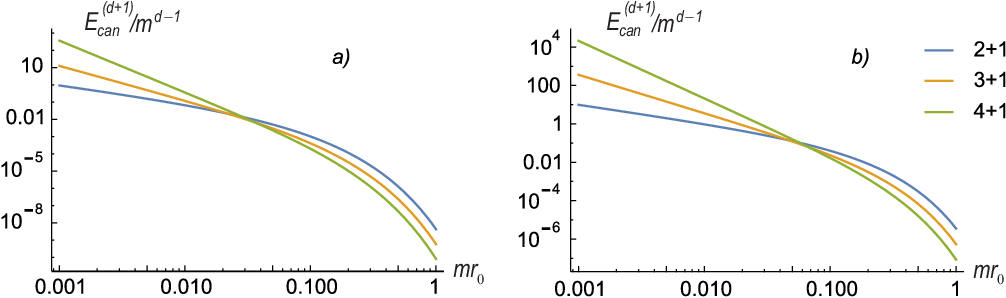}\;
   
    \caption{The induced dimensionless vacuum energy $E_{can}^{d+1}/m^{d-1}$ within the plane transverse to the tube in $(d+1)$-dimensional space-time as the function of the thicknesses of the magnetic impenetrable tube $mr_0$ for the cases of \textit{a}) Dirichlet and \textit{b}) Neumann boundary conditions.}
    \label{Fig4}
\end{figure}

Using the results of \cite{Gorkavenko:2013rsa,Gorkavenko:2022xtv}, we reconstruct the dependence of the induced dimensionless vacuum energy $E^{(d+1)}_{can}/m^{d-1}$ on the tube thickness for the Dirichlet and Neumann cases, see Fig.\ref{Fig4}, and compare it with the maximal value of $E^{(d+1)}_\xi/m^{d-1}$ attained at some $\theta_{\mathrm{max}}$ in the range $-\pi/2 < \theta < 0$, see Table~\ref{Table}. The value of $E^{(d+1)}_{\xi}/m^{d-1}$ at $\theta_{\mathrm{max}}$ lies between those for the Dirichlet and Neumann boundary conditions. As seen from Table~\ref{Table}, for certain values of the boundary condition parameter $-\pi/2 < \theta < 0$, the quantity $E^{(d+1)}_\xi/m^{d-1}$ is of the same order of magnitude, or even larger, than $E^{(d+1)}_{can}/m^{d-1}$ for the same $\theta$.  It has been shown that this statement is true for cases of different dimensions of space-time. 

 {We would also like to note that Fig.\ref{Fig4} shows the dependence of the canonical contribution to the induced energy, $E^{(d+1)}_{can}$, on the thickness of the magnetic tube for a fixed boundary-condition parameter. 
One can see that $E^{(d+1)}_{can}$ increases rapidly  as $mr_0$ decreases, and the same behavior is exhibited by the $E^{(d+1)}_{\xi}$
 contribution in Fig.\ref{Fig3}.  However, Fig.\ref{Fig4} additionally allows one to see that $E^{(d+1)}_{can}$ increases with increasing space dimension $d$ only for small tube radius, $mr_0 < mr_{c}$, and decreases for large tube radius, $mr_0 > mr_{c}$. For the cases $d = 2,3,4$, we have $mr_{c} \approx 0.05$, and it grows slowly as the spatial dimension increases. The same behavior can be observed from a detailed analysis of Fig.\ref{Fig3} and Table~\ref{Table} for the $E^{(d+1)}_{\xi}$ contribution to the induced energy. }

\begin{table}[t]
\centering
\begin{tabular}{c|c|c|c|c|c}
\hline  $m r_0$ & 1 & $1/2$ & $10^{-1}$ & $10^{-2}$ & $10^{-3}$\\
\hline $E^{(2+1)}_{can,D}/m\phantom{{}^2}$ & $4.36\cdot10^{-9}\phantom{{}^2}$
& $1.30\cdot10^{-6}$ & $1.04\cdot 10^{-3}$ &
$0.066$ & $0.933$\\
\hline $E^{(2+1)}_{can,N}/ m$ &
$3.52\cdot10^{-6}$ & $3.00\cdot10^{-4}$  & $3.96\cdot10^{-2}$ & $0.941$
 & 9.80\\
\hline
 $E^{(2+1)}_{\xi,max}/ m$ & 
$4.55\cdot10^{-6}$ & $3.78\cdot10^{-4}$ & $4.58\cdot10^{-2}$ &
$1.00$ & $10.75$\\
\hline
\end{tabular}\vspace{0.5em}
\begin{tabular}{c|c|c|c|c|c}
\hline $E^{(3+1)}_{can,D}/m^2$ &  $5.94\cdot10^{-10}$
& $2.41\cdot10^{-7}$ & $4.16\cdot 10^{-4}$ &
$0.119$ & $12.7$\\
\hline $E^{(3+1)}_{can,N}/ m^2$ & 
$5.40\cdot 10^{-7}$ & $6.71\cdot 10^{-5}$ & $2.31\cdot 10^{-2}$ & 3.55  & 359\\
\hline
 $E^{(3+1)}_{\xi,max}/ m^2$ &
$7.1\cdot10^{-7}$ & $8.40\cdot10^{-5}$ & $2.87\cdot10^{-2}$ &
$4.25$ & $422$\\
\hline
\end{tabular}\vspace{0.5em}
\begin{tabular}{c|c|c|c|c|c}
\hline $E^{(4+1)}_{can,D}/m^3$ &  $8.35\cdot10^{-11}$
& $4.73\cdot10^{-8}$ & $2.00\cdot 10^{-4}$ &
$0.365$ & $371$\\
\hline $E^{(4+1)}_{can,N}/ m^3$ &  $8.53\cdot 10^{-8}$
 & $1.60\cdot 10^{-5}$ & $1.66\cdot10^{-2}$ & $21.0$  & 21040 \\
\hline
 $E^{(4+1)}_{\xi,max}/ m^3$ &  
$1.2\cdot10^{-7}$ & $1.20\cdot10^{-5}$ & $2.13\cdot10^{-2}$ &
$26.2$ & $26300$\\
\hline
\end{tabular}
\caption{Values of the dimensionless induced vacuum energy within the plane transverse to the tube in $(d+1)$-dimensional space-time. $E^{(d+1)}_{can,D(N)}/ m^{d-1}$ corresponds to the case of Dirichlet (Neumann) boundary condition, $E^{(d+1)}_{\xi,max}/ m^{d-1}$ corresponds to the maximum value of $E^{(2+1)}_\xi$ at $-\pi/2 < \theta < 0$.} \label{Table}
\end{table}

\section{ {Asymptotic behavior of the induced $\xi$-dependent energy for thin and thick tube limits}}

 {
In this section, we will closely examine the dependence of the induced vacuum energy $E_\xi^{(d+1)}$ on the thickness of the tube, namely on the parameter $mr_0$.}

 {The study of the asymptotic behavior of the $\xi$-dependent induced vacuum energy in the limits of very thin and very thick magnetic tubes is motivated by the need to clarify how the coupling to space-time curvature manifests itself in extreme physical regimes. In the thin-tube limit, the magnetic tube effectively reduces to a zero-thickness string, while the thick-tube limit probes the suppression of vacuum effects under the increasing of the tube radius $r_0$. 
 On the other hand, the thickness of a topological defect can be estimated from the approximate condition $m_H r_0 \sim  1$, where $m_H$ is the scale of spontaneous symmetry breaking (or the mass of the corresponding Higgs field) responsible for the formation of the topological defect in astrophysical or condensed-matter systems.
This relation implies $mr_0 \sim m/m_H$.
Therefore, the consideration of thin and thick tube limits effectively corresponds to studying vacuum polarization effects for light and heavy scalar fields relative to the symmetry-breaking scale.}


For $mr_0\ll 1$, we get from \eqref{4s5}
\begin{equation}\label{4s5ap}
\frac{E^{(d+1)}_{\xi}}{m^{d-1}}
  =\frac{C(F,d,\theta)}{x_0^{d-1}},\qquad C(F,d,\theta)=2\frac{(4\pi)^{1-d/2}}{\Gamma\left(\frac{d-2}2\right)}\int\limits_{0}^\infty
dv\, v^{d-2} \,\mathcal{D}_\xi(F,\theta,v).
\end{equation}
Unfortunately, the coefficient $C(F,d,\theta)$ cannot be written in analytical form. But numerical computations show that it is a decreasing function of $d$, which can be roughly approximated for $d\leq 8$ as 
\begin{equation}
    C= e^{-a-b\, d },
\end{equation}
where $a$, $b$ are positive coefficients of $F$ and $\theta$ parameters. Then at $mr_0\ll 1$  we get
 {\begin{equation}\label{4s5ap1}
\frac{E^{(d+1)}_{\xi}}{m^{d-1}}=e^{\ln x_0-a}e^{-d(b+\ln x_0)}
  =\frac{\beta}{(\gamma x_0)^{d-1}}, \quad mr_0\ll 1, 
\end{equation}
where $\beta=e^{-(a+b)}$, $\gamma=e^{b}$.} Such a simple representation makes it possible, at a qualitative level, to understand the dependence (obtained numerically) of the induced vacuum energy on the tube thickness and on the space dimension. 
By analyzing the function in the exponent, one can see that for a magnetic tube of small thickness\footnote{If we put it more precisely, the function $C(d)$ has the form $C= e^{\alpha+\beta\, d + \gamma/d +\delta\gamma/d^2}$ and $mr_c=e^{\beta - {2 \delta}/{d^3} - {\gamma}/{d^2}}$. For example, for $F=1/2$ and $\theta=-1$, the coefficients are $\alpha=-8$, $\beta=-1.78$, $\gamma=20.48$, and $\delta=-14.2$, which, in turn, give $mr_c=0.05$ for $d=3$ and $mr_c=0.073$ for $d=4$.}, $x_0<mr_c$, $mr_c=e^{-b}$, the induced vacuum energy increases as the space dimension $d$
 grows, whereas for a sufficiently large tube thickness, the induced vacuum energy instead decreases with increasing $d$. For example, for $F=1/2$ and $\theta=-1$, the coefficients are $a=0.55$ and $b=2.48$, which, in turn, give $mr_c=0.084$  {and $\beta=0.048$, $\gamma=11.94$.}

For $mr_0\gg1$, we get from \eqref{4s5} 
\begin{equation}\label{4s5AssLarge}
E^{(d+1)}_{\xi}
  =\frac2{r_0^{d-1}}\frac{(4\pi)^{1-d/2}}{\Gamma\left(\frac{d-2}2\right)}\int\limits_{x_0}^\infty
dv\, v^2 \left(v^2-x_0^2\right)^{\frac{d-4}2}\,\mathcal{D}^{as}_\xi(F,\theta,v),
\end{equation}
where $\mathcal{D}^{as}_\xi$ is the asymptotic form of the function $\mathcal{D}_\xi$ \eqref{dddop4} at large $mr_0$. Numerical computations show that it can be 
approximated as
\begin{equation}
    \mathcal{D}^{as}_\xi=\frac{e^{-\bar a -\bar b \,x_0}}{x_0}, \quad x_0 \gg 1,
\end{equation}
where $\bar a$, $\bar b$ are positive coefficients of $F$ and $\theta$ parameters.
By substituting the obtained asymptotic expression into \eqref{4s5AssLarge}, one can carry out the integration explicitly and obtain
\begin{equation}
 \frac{E^{(d+1)}_{\xi}}{m^{d-1}}=   \frac{2 \bar b}{e^{\bar a} (2 \pi \bar b\, x_0)^{\frac{d - 1}2}} K_\frac{d-1}{2}(\bar b\,  x_0)  {\approx  \frac{\bar b\, e^{-\bar a }}{(2 \pi )^{\frac{d}{2}-1}} \frac{e^{- \bar b\, x_0}}{(\bar b x_0)^{\frac{d}{2}}},} \quad mr_0\gg 1,
\end{equation}
where $K_\mu(u)$ is the Macdonald function. 
For example, for $F=1/2$ and $\theta=-1$, the coefficients are $\bar a= 5.22$ and $\bar b=7.23$.

 {To conclude this section, it is important to note that variations in the tube radius in physical systems may be accompanied by changes in the surface properties of the tube and the corresponding boundary parameter $\theta$, see, e.g., \cite{KaneMele1997,Bordag:2005by}.}

\section{Summary} \vspace{-0.5em}

 {In this paper, we considered the possibility of the manifestation of coupling $\xi$ of the scalar field’s interaction with the space-time curvature in careful measurements of vacuum polarization effects in flat space-time.}

 {A nonzero value of the coupling $\xi$ of the scalar field’s interaction with the space-time curvature modifies the effective mass of particles, which affects particle creation and decay processes, influences the inflationary stage of the Universe’s evolution, the stability of the electroweak vacuum, and the spectrum of Hawking radiation from black holes for scalar fields. At the same time which exact value of the coupling constant is realized in nature remains an open question. An experimental determination is hindered by the typically negligible value of the Ricci scalar $R$ for most cosmological and astrophysical problems. In fact, the contribution of the curvature coupling constant becomes significant when the scalar field is in regions with large curvature, such as in the early Universe (during inflationary or radiation-dominated epochs) or near strong gravitational sources (black holes, neutron stars). Therefore, studies of the possible manifestation of coupling $\xi$ in flat space-time are of considerable interest.}

In flat $(d+1)$-dimensional space-time, we consider the vacuum polarization of a quantized charged scalar field in the presence of a magnetic topological defect, modeled as an impenetrable tube of finite thickness carrying magnetic flux. 
The induced vacuum energy is studied with particular attention to its dependence on the coupling parameter $(\xi)$, which characterizes the interaction of the scalar field with space-time curvature. The scalar matter field is assumed to be excluded from the tube interior and to satisfy  {general} Robin-type boundary conditions \eqref{Robin} on its surface. These boundary conditions are specified by a single parameter, $-\pi/2 \leq \theta < \pi/2$. The special cases $\theta=0$ and $\theta=-\pi/2$ correspond to Dirichlet and Neumann boundary conditions, respectively. 

We get general relations for the induced vacuum energy within the plane transverse to the tube in $(d+1)$-dimensional space-time \eqref{dddop5}.
The imposed impenetrability of the matter field into the magnetic field region naturally gives rise to the Aharonov–Bohm effect. As a result,  the induced vacuum energy exhibits a periodic dependence on the magnetic flux inside the tube and depends only on the fractional part of the magnetic flux  $F$ \eqref{a29a1}. When the magnetic flux $F$ assumes an integer value, the effects of vacuum polarization disappear.

Our analysis demonstrates that, in flat space-time, the  induced vacuum energy within the plane transverse to the
tube $E^{(d+1)}=E^{(d+1)}_{can}+(1/4-\xi)E^{(d+1)}_\xi$ is independent of the coupling 
$\xi$ of the scalar field's interaction with the space-time curvature only in the special cases of Dirichlet and Neumann boundary conditions  {$(E^{(d+1)}_\xi=0)$}. 
But for general Robin boundary conditions, the induced vacuum energy in flat space-time depends on the coupling $\xi$  {$(E^{(d+1)}_\xi\neq 0)$} and exhibits a nontrivial dependence on the parameter of the boundary condition $\theta$.

We computed $E^{(d+1)}_\xi$, which determines the dependence of the
total induced vacuum energy on the coupling $\xi$ for the cases of $d=2,3,4$ and negative values of boundary parameter  $-\pi/2 \leq \theta \leq 0$, see Figs.\ref{Fig1} and \ref{Fig3}. 
As one can see, the vacuum effects increase strongly as the tube radius decreases. 
It is also worth noting that as the tube thickness decreases, the value of the boundary condition parameter at which the function reaches its maximum shifts toward the Neumann boundary condition, $\theta = -\pi/2$. Importantly, the induced energy increases with increasing the space-time dimensionality and for sufficiently small transverse sizes of the topological defect  {$mr_0<mr_{c}$, but decreases with increasing dimensionality at large $mr_0>mr_{c}$, where $mr_{c}\approx 0.05$ for $d=2,3,4$ and grows slowly as the spatial dimension increases.}

We reconstruct the dependence of the induced dimensionless vacuum energy $E^{(d+1)}_{can}/m^{d-1}$ on the tube thickness for the Dirichlet and Neumann cases, see Fig.\ref{Fig4}, and compare it with the maximal value of $E^{(d+1)}_\xi/m^{d-1}$ attained at some $\theta_{\mathrm{max}}$, see Table \ref{Table}. We conclude that for certain values of the boundary condition parameter $-\pi/2 < \theta < 0$, the quantity $E^{(d+1)}_\xi/m^{d-1}$ is of the same order of magnitude, or even larger, than $E^{(d+1)}_{can}/m^{d-1}$ for the same $\theta$. This means that the $E^{(d+1)}_\xi$ plays an important role in determining the induced energy, and the dependence of  {the induced vacuum energy \eqref{dddop5}} on the parameter $\xi$ is significant.

We conclude that careful measurements of vacuum polarization effects in flat space-time may, in principle, provide an independent way to probe the $\xi$ coupling that determines the interaction of a scalar field with space-time curvature.   {Since the induced vacuum energy effect is strongest for tubes with $m r_0 \ll 1$, efforts to observe the $E_\xi$ contribution should focus on satisfying this requirement. Additionally, it is essential to establish conditions corresponding to the value of the boundary parameter $\theta$ that maximizes $E_\xi$.}



 { It should be noted that the results obtained here for the physically interesting $(3+1)$-dimensional space–time, as well as for higher-dimensional spaces, were derived by performing the corresponding integration of the induced vacuum energy expression in $(2+1)$-dimensional space–time, see \eqref{dddop5}. The results for the induced vacuum energy generated by a magnetic tube with a general Robin boundary condition in $(2+1)$-dimensional space–time were first obtained in \cite{Gorkavenko:2024vuy}. In that paper, the induced vacuum energy was studied only for a fixed thickness of the magnetic tube (ring), $mr_0=1/100$, and for various negative values of the Robin boundary parameter $\theta$. In the present paper, we compute the induced energy for an arbitrary thickness of the magnetic tube in $(2+1)$-dimensional space–time, which makes it possible to carry out the integration procedure used in \cite{Gorkavenko:2024vuy} and thus extend the previous results to space–times of arbitrary dimensionality $d$ and arbitrary values of the transverse size of the magnetic tube $mr_0$.}

When the boundary condition parameter $\theta$ takes positive values, the problem becomes particularly interesting and requires further thorough analysis. In this case, bound-state solutions of the Fock–Klein–Gordon equation may contribute to vacuum polarization, which can result in nontrivial discontinuous dependence of vacuum effects on the tube thickness and $\theta$, similar to the behavior observed for the induced magnetic flux \cite{Sitenko:2022gha}.


\section*{Acknowledgments}

The work of V.M.G. and M.S.Ts. was supported by Department of target training of
Taras Shevchenko National University of Kyiv and the NAS of Ukraine,
grant No.6$\Phi$-2024. 

\newpage

\setcounter{equation}{0}
\renewcommand{\theequation}{A.\arabic{equation}}
\section*{Appendix A: Evaluating of $E_\xi^{(d+1)}$}\label{AppendixA}

Let us start from the general definition in \eqref{4s5}. Let us switch to the spherical coordinate system in a 
$(d-2)$-dimensional momentum space
\begin{equation}\label{A1}
    \int\limits_{-\infty}^\infty d^{d-2}{\bf
p}=\int d\Omega_{d-2}\int_0^\infty p^{d-3} dp, \quad \int d\Omega_{d-2}=2\frac{\pi^{\frac{d}2-1}}{\Gamma(\frac{d}2-1)}.
\end{equation}
After transfer to dimensionless variables $x=mr$, $x_0=mr_0$, $z=kr$, $\alpha=pr$ one gets
\begin{equation}\label{A2}
E^{(d+1)}_{\xi}=(2\pi)^{2-d}\int d\Omega_{d-2}\int\limits_{mr_0}^\infty dx \,x \Delta_x \int_0^\infty \alpha^{d-3} d\alpha \int\limits_0^\infty
  dz\,z\,\frac{m^{d-1}}{x^{d-1}}\frac{ G(\theta,z,z\frac{x_0}{x},F)}{\sqrt{\alpha^2+z^2+x^2}}.
\end{equation}
After changing of variables $\alpha=x u^\beta$ we get
\begin{equation}\label{A3}
E^{(d+1)}_{\xi}=(2\pi)^{2-d}\,\beta \int d\Omega_{d-2}\int\limits_0^\infty du\, u^{\beta(d-2)-1}\int\limits_{mr_0}^\infty dx \,x \Delta_x \int\limits_0^\infty
  dz\,z\, \frac{m^{d-1}}{x}\frac{ G(\theta,z,z\frac{x_0}{x},F)}{\sqrt{z^2+x^2(1+u^{2\beta})}}.
\end{equation}
Taking $\tilde x=x\sqrt{1+u^{2\beta}}$, we get
\begin{multline}\label{A4}
E^{(d+1)}_{\xi}=(2\pi)^{2-d}\,m^{d-1}\beta \int d\Omega_{d-2}\int\limits_0^\infty du\, u^{\beta(d-2)-1} (1+u^{2\beta})^{\frac12}\times \\ \int\limits_{x_0\sqrt{1+u^{2\beta}}}^\infty d\tilde x \,\tilde x \Delta_{\tilde x} \int\limits_0^\infty
  dz\,z\, \frac{1}{\tilde x}\frac{ G(\theta,z,z\frac{x_0\sqrt{1+u^{2\beta}}}{x},F)}{\sqrt{z^2+\tilde x^2}}=\\
  (2\pi)^{2-d}\,m^{d-1}\beta \int d\Omega_{d-2}\int\limits_0^\infty du\, u^{\beta(d-2)-1} (1+u^{2\beta})^{\frac12} \mathcal{D}_{\xi}\left(\theta,x_0\sqrt{1+u^{2\beta}}\right),
\end{multline}
where $\mathcal{D}_{\xi}$ was defined in \eqref{dddop4}.  Putting $\beta=(d-2)^{-1}$, taking into account the result of integration over angular variables \eqref{A1}, and the definition of $E^{(2+1)}_{\xi}$ \eqref{dop3}, we get 
\begin{equation}\label{A5}
E^{(d+1)}_{\xi}
  =m^{d-1}\frac{(4\pi)^{1-d/2}}{\Gamma(d/2)}\int\limits_0^\infty
du\,\sqrt{1+u^{2/(d-2)}}\,
\mathcal{D}_{\xi}\left(\theta,x_0\sqrt{1+u^{2/(d-2)}}\right).
\end{equation}
Finally, changing the integration variable $v=x_0\sqrt{1+u^{2/(d-2)}}$, we get the relation \eqref{4s5}.

\newpage
\bibliographystyle{JHEP}
\bibliography{bibliography}

\end{document}